\documentclass{article}
\usepackage[utf8]{inputenc}
\usepackage{graphicx,amssymb}
  \newcommand{\bq}{\begin{quote}} \newcommand{\eq}{\end{quote}}     \newcommand{\M}{\mid\mid} \newtheorem{Th}{Theorem}   \newtheorem{ax}{Axiom}   \newtheorem{lm}{Lemma}  \newtheorem{df}{Definition}     \newtheorem{pr}{Proposition}  \newtheorem{cl}{Corollary}   \newtheorem{re}{Remark}     \newtheorem{as}{Assumption}   \newtheorem{wg}{Wild Guess} \newtheorem{ex}{Example} \newcommand{\bth}{\begin{Th}\hspace{-5pt}{\bf .} \ }  \newcommand{\Eth}{\end{Th}} \newcommand{\bax}{\begin{ax}\hspace{-5pt}{\bf .} \ }  \newcommand{\eax}{\end{ax}} \newcommand{\blm}{\begin{lm}\hspace{-5pt}{\bf .} \ } \newcommand{\elm}{\end{lm}} \newcommand{\bdf}{\begin{df}\hspace{-5pt}{\bf .} \ }    \newcommand{\edf}{\end{df}}  \newcommand{\bpr}{\begin{pr}\hspace{-5pt}{\bf .} \ }  \newcommand{\epr}{\end{pr}} \newcommand{\bcl}{\begin{cl}\hspace{-5pt}{\bf .} \ }  \newcommand{\ecl}{\end{cl}} \newcommand{\bre}{\begin{re}\hspace{-5pt}{\bf .} \ } \newcommand{\ere}{\end{re}} \newcommand{\bas}{\begin{as}\hspace{-5pt}{\bf .} \ } \newcommand{\eas}{\end{as}} \newcommand{\bwg}{\begin{wg}\hspace{-5pt}{\bf .} \ } \newcommand{\ewg}{\end{wg}} \newcommand{\bex}{\begin{ex}\hspace{-5pt}{\bf .} \ }   \newcommand{\eex}{\end{ex}}   \newcommand{\bit}{\begin{itemize}} \newcommand{\eit}{\end{itemize}\par\noindent} \newcommand{\beq}{\begin{equation}}  \newcommand{\eeq}{\end{equation}\par\noindent} \newcommand{\beqa}{\begin{eqnarray*}} \newcommand{\eeqa}{\end{eqnarray*}\par\noindent} \newcommand{\beqn}{\begin{eqnarray}}   \newcommand{\eeqn}{\end{eqnarray}\par\noindent}
\newcommand{\diraco}{\mid 0 \rangle}
\newcommand{\dirac}{\mid 1 \rangle}
\newcommand{\Odiraco}{\overline{\mid 0 \rangle}}
\newcommand{\Odirac}{\overline{\mid 1 \rangle}}

\title{{\bf The logic of quantum programs}}
\author{Alexandru Baltag\footnote{Oxford University Computing Laboratory}
 $\,$ and Sonja Smets\footnote{Vrije Universiteit Brussel, Flanders Fund for Scientific Research Post-Doc}}
\date{}
\begin{document}
\maketitle

\begin{abstract}
\par\noindent
    We present a logical calculus for reasoning about information flow in quantum programs. In particular we introduce a dynamic logic that is capable of dealing with quantum measurements, unitary evolutions and entanglements in compound quantum systems. We give a syntax and a relational semantics in which we abstract away from phases and probabilities. We present a sound proof system for this logic, and we show how to characterize by logical means various forms of entanglement (e.g. the Bell states) and various linear operators. As an example we sketch an analysis of the teleportation protocol.
\end{abstract}
\section{Introduction}
In this paper we elaborate on the ideas presented in \cite{Baltag,BaltagSmets,Smets} and give a full-fledged {\it dynamic Logic for Quantum Programs LQP}. It is well-known that {\it PDL} (Propositional Dynamic Logic) and its fragment Hoare Logic are among the main logical formalisms used in {\it program verification} for classical programs, i.e. in checking that a given (classical) program meets the required specification. It is natural to ask for a {\it quantum} version of {\it PDL}, to be used in the verification of quantum programs. In our past work \cite{BaltagSmets}, we presented several such logical systems, and later extending this system into a dynamic logic $LQA$ of {\it quantum actions} (i.e. compositions of measurements and unitary evolutions). In this paper, we extend $LQA$ into a logic for {\it compound} quantum systems. We present a self-contained version of $LQP$ such that no knowledge of $LQA$ or $LQM$ is necessary to understand the basic concepts. Note the difference between our logic and the approach with a similar name in \cite{Brunet}: our dynamic logic goes much further in capturing essential properties of quantum systems and quantum programs, as well as in recovering the ideas of traditional quantum logic \cite{DC,G}.

%%%%%%%%%%%%%%%%%%%%%%%%%%%%%%%%%%%%%%%
%\cite{BaltagSmets, LICS,Portugal}.
%Several logical systems have been proposed:
%\cite{BaltagSmets} we focused on single quantum systems and presented two equivalent {\it
%\cite{Baltag, Smets}
%in \cite{BaltagSmets},
%\cite{Daniel1,Daniel2,Faure,Amira,LogicOfDynamics,Compoundness,Quantale,
%SasakiIsNot,CS,MalPaper} \cite{Hardegree1,Hardegree2}
% \cite{Beltrametti,OnCausation}.}

%%%%%%%%%%%%%%%%%%%%%%%%%%%%%%%%%%%%%%%
\section{Quantum Frames}
%%%%%%%%%%%%%%%%%%%%%%%%%%%%%%%%%%%%%%%
In this section we introduce quantum frames for single quantum systems and quantum frames for compound quantum systems; in the later case we restrict our attention ro $n$ compound qubits.

\subsection{Single System Quantum Frames}
%%%%%%%%%%%%%%%%%%%%

A {\it modal frame} is a set of {\it states}, together with
a family of {\it binary relations} between states. A (generalized) {\it PDL frame}
is a modal frame $(\Sigma, \{\stackrel{S?}{\to}\}_{S \in {\cal L}},
\{ \stackrel{a}{\to} \}_{a \in {\cal A}})$,
in which the relations on the set of states $\Sigma$ are of two types:
the first, called {\it tests} and denoted by $S?$, are labelled with
subsets $S$ of $\Sigma$, coming from a given family ${\cal L}\subseteq
{\mathcal P}(\Sigma)$ of sets, called {\it testable properties}; the
others, called {\it actions}, are labelled with action labels $a$ from
a given set ${\cal A}$.

Given a $PDL$ frame, there exists a standard way
to give a semantics to the usual language of {\it propositional
dynamic logic}. Classical $PDL$ can be considered as a special case
of such a logic, in which tests are given by {\it classical tests}:
$s \stackrel{S?}{\to} t$ if and only if $s = t\in S$.
Observe that {\it classical tests, if executable, do not change the
current state}.

In the context of quantum systems, a natural idea is to replace
classical tests by ``quantum tests'', given by {\it quantum
measurements} of a given property. Such tests will obviously change
the
state of the system. To model them, we introduce a special kind of $PDL$ frames: {\it quantum
frames}. The ``tests'' are essentially given by {\it projectors} in a
Hilbert space. In  \cite{BaltagSmets}, we considered $PDL$ with the above-mentioned standard
semantics, having the same clauses in the classical case, but
interpreted in quantum frames. What we obtained is a {\it quantum PDL}, whose
negation-free
part with dynamic modalities for quantum tests is equivalent to what is traditionally called ``(orthomodular)
quantum logic'' \cite{DC,G}. In this paper, we extend the syntax of this logic to
deal with unitary evolutions, entanglements and some quantum protocols.

\smallskip\par\noindent
\textbf{Definition 1.} {\it (Quantum Frames)}
\par\noindent
Given a Hilbert space ${\cal H}$, the following steps construct a
{\it Quantum (PDL) Frame} $$ {\Sigma({\cal H})} :=  {(\Sigma,
\{\stackrel{S?}{\to}\}_{S \in {\cal L}}, \{ \stackrel{U}{\to} \}_{U \in {\cal U}})}$$
\begin{enumerate}
\item Let $\Sigma$ be the set of {\it one dimensional subspaces} of
${\cal H}$, called the set of {\it states}. We denote a state
$s=\overline{x}$ of ${\cal H}$ using any of the non-zero vectors $x
\in {\cal H}$ that generate them. Note that any two vectors
that differ only in {\it phase} (i.e. $x=\lambda y$, with $\lambda\in
C$ with $|\lambda|=1$) will generate the same state
$\overline{x}=\overline{y}\in \Sigma$.
\item Call two states $s$ and $t$ in $\Sigma$ {\it orthogonal}, and
write $s \perp t$, if and only if $\forall x\in s$ and $\forall y\in t$: $x$ is orthogonal to $y$, i.e. if
$ \langle x
\mid y \rangle = 0$. Or, equivalently, we can state that $s \perp t$ if and only if $\exists x  \in s,  y \in t$ with $x  \not = 0$, $y \not
= 0$ and $ \langle x \mid y \rangle = 0$.
We put $S^{\perp} := \{t \in \Sigma \mid t \perp s \mbox{ for \, all }s
\in S \}$; and we denote by $\overline{S}=S^{\perp \perp} := (S^{\perp})^{\perp}$ the
biorthogonal closure of $S$. In particular, for a singleton
$\{x\}$, we just write $\overline{x}$ for $\overline{\{x\}}$, which
agrees with the notation $\overline{x}$ used above to denote the state
generated by $x$.
\item A set of states $S \subseteq \Sigma$ is called a {\it (quantum) testable
property} iff it is
{\it biorthogonally
closed}, i.e. if $\overline{S} = S$. (Note that $S \subseteq \overline{S}$ is always the case.)
We denote by
${\cal L}\subseteq P(\Sigma)$ the family of all quantum testable
properties. All the {\it other} sets $S\in P(\Sigma)\setminus {\cal
L}$ are called {\it non-testable properties}.
\item There is a natural bijective correspondence between the family
${\cal L}$ of all testable
properties and the family ${\cal W}$ of all {\it closed linear
subspaces} $W$ of ${\cal H}$, bijection given by
$S \, \, \mapsto \, \, W_S=:\bigcup S$. Observe that, under this
correspondence,
the image of the biorthogonal closure $\overline{S}$ of any arbitrary set
$S\subseteq \Sigma$ is the closed linear subspace $\overline{\bigcup
S}\subseteq {\cal H}$ generated by the union $\bigcup S$ of all states
in $S$.
\item For each testable property $S\in {\cal L}$, there exists a
partial map $S?$ on $\Sigma$, called a {\it quantum test}. If
$W= W_S=\bigcup S$ is the corresponding subspace of ${\cal H}$, then
the quantum test is the map induced on states by the {\it projector}
$P_W$ onto the subspace $W$. In other words, it's given by:
\begin{eqnarray*}
S?(\overline{x}) & := & \overline{P_W(x)} \in \Sigma\, , \,    \mbox{
if }\overline{x} \not \in S^\perp \,  ( \mbox{ i.e.  if } P_W(x) \not
= 0) \\
S?(\overline{x}) & := & \mbox{ undefined }\, , \, \mbox{ otherwise }.
\end{eqnarray*}
We denote by $\stackrel{S?}{\to} \subseteq \Sigma \times \Sigma$ the
binary relation corresponding to the partial map $S?$, i.e. given by:
$s \stackrel{S?}{\to} t$ if and only if $S?(s) = t$. So we have
{\it a family of binary relations indexed by the testable properties}
$S \in {\cal L}$.

\item For each unitary transformation $U$ on ${\cal H}$, consider the
corresponding binary relation $\stackrel{U}{\to} \subseteq \Sigma
\times \Sigma$, given by:  $s \stackrel{U}{\to} t$ if and only if
$ U (x)  =  y $ for some  non-zero vectors $x \in s, y  \in t$. So we obtain
{\it a family of binary relations indexed by the unitary
transformations} $U \in {\cal U}$ (where ${\cal U}$ is the set of
unitary transformations on ${\cal H}$).
\end{enumerate}

\par\noindent
So a quantum frame is just a $PDL$ frame built on top of a given Hilbert
space ${\cal H}$, using projectors as ``tests'' and unitary evolutions
as ``actions''.
Our notion of ``state'' in this paper
is closely connected to the way quantum logicians approach quantum
systems, i.e., contrary to identifying states with unitary vectors (as customary in quantum computation), we took them to be {\it one dimensional subspaces} generated by these vectors. This imposes some limits to our approach, mainly that we will not be able
to express {\it phase}-related properties. While it is possible to build up a quantum frame starting from unitary vectors as the states, the resulting logical system will be much more complex\footnote{It would require the introduction of a propositional {\it tensor} operator.}, and so we do not elaborate on it in this paper.

%\pagebreak
\smallskip\par\noindent
{\bf Operators on states, adjoints and generalized tests.}
To generalize our notations introduced earlier, observe that every
{\it linear operator} $F: {\cal H} \to {\cal H}$ induces
a partial map $F: {\Sigma} \to {\Sigma}$ on states (i.e. subspaces), given by
$F(\overline{x}) = \overline{F(x)} $. (Note that {\it linearity}
ensures that this map on states is well-defined.) In particular, every
map $F:\Sigma\to \Sigma$ obtained in this way has an {\it adjoint}
$F^{\dagger}:\Sigma\to \Sigma$, defined as the map on states induced by
the adjoint (``Hermitian conjugate'') of the linear operator $F$ on
${\cal H}$. Observe that, for unitary transformations $U$, the adjoint
is the inverse: $U^{\dagger}=U^{\mbox{-1}}$.
Also, one can
naturally generalize {\it quantum tests} to arbitrary, possibly {\it
non-testable properties}, $S\subseteq \Sigma$, by putting: $S ? :=
\overline{S} ?$. So we identify a test of a ``non-testable'' property $S$
with the quantum test of its biorthogonal closure. Observe that $S?^{\dagger}=S?$
(since projectors are self-adjoint).

\smallskip\par\noindent\textbf{Definition 2.} {\it(Non-orthogonality, or Measurement Relation)}.
\par\noindent
For all $s,t \in \Sigma$, let $s \to t$ if and only if $s
\stackrel{S?}{\to} t$ for some property $S \in {\cal L}$.
In other words, $s\to t$ means that one can reach state $t$ by doing {\it
some measurement} on state $s$.
An important observation is that {\it the measurement relation is the
same as non-orthogonality}: $s\to t$ iff $s\not\perp t$. The non-orthogonality relation has indeed been used to
 introduce an accessibility relation in the orthoframe semantics within quantum logic \cite{G}.

\smallskip\par\noindent\textbf{Definition 3.} {\it(Dynamic Modalities and Measurement Modalities)}
\par\noindent
{\it For
any property $T\subseteq \Sigma$
and any partial map $F: \Sigma \to \Sigma$ induced on states by a linear operator $F$, let
$[F] T := F^{\mbox{-1}}(T) = \{s\in \Sigma : F(s) \in T \mbox{ if defined }\}$ and
$\langle F \rangle T :=  \Sigma \backslash ([F] (\Sigma \backslash
F))$. Similarly, put
$\Box T := \{s \in \Sigma : \forall t (s \to t \Rightarrow t \in T)\} $ and $\, \Diamond T := \Sigma \backslash (\Box (\Sigma \backslash T))$.
}
%%%%%%%%%%%%%%%%%
\smallskip\par\noindent
Observe that $[F]T$ expresses the {\it weakest precondition} for the
``program'' $F$ and post-condition $T$. In particular,
$[S?]T$  expresses the weakest precondition ensuring the satisfaction
of property $T$ in any state after the system passes a quantum test of
property $S$. Similarly, $\langle S? \rangle T$ means that one can
perform a quantum test of property $S$ on the current state, ending up
in a state having property $T$.
$\Box T$ means that property $T$ will hold after {\it any} measurement
(quantum test) performed on the current state. Finally, $\Diamond T$
means that property $T$ is {\it potentially satisfied}, in the sense
that one can do some quantum test to reach a state with property $T$.

\smallskip\par\noindent
{\bf Lemma 1.} {\it For every property $S\subseteq \Sigma$, we have
$S^{\perp}=[S?]\emptyset=\Sigma\setminus \Diamond S$ and
$\overline{S}= \Box\Diamond S$.
}

\smallskip\par\noindent
{\bf Proposition 1.} {\it For every property $S\subseteq \Sigma$, if $T\in {\cal L}$
(i.e. is testable), then $\Box S, S^{\perp}, [S?]T\in {\cal L}$ (are
testable), and more generally $[F] T\in {\cal L}$, for every (map on states induced by a) linear operator $F$.}

\smallskip\par\noindent
{\bf Proposition 2.} {\it (Testable Properties) A property $S \subseteq \Sigma$ is testable if and only if any of the
following equivalent conditions hold: (1) $S = \overline{S}$; (2) $S = \Box \Diamond P$; (3) $\exists T \in \Sigma \mbox{ \, such \, that } P = T^{\perp}$; (4)  $\exists T \in \Sigma \mbox{\, such \, that } P = \Box T$. The family ${\cal L}$ of testable properties is a complete lattice with
respect to inclusion, having as its meet set-intersection $S\cap T$,
and as its join the
biorthogonal closure of set-union $S\sqcup T:= \overline{S\cup T} \,,$
called the quantum join of $S$ and $T$. For every state $s \in \Sigma$, the singleton $\{s\} \in {\cal L}$ is testable. For any arbitrary property
$S\subseteq \Sigma$, we have $\overline{S}=\bigsqcup\{\{s\}: s\in S\}=
\bigcap\{T\in {\cal L}: S\subseteq T\}$, so the biorthogonal closure of
$S$ is the strongest testable property implied by (the property) $S$.
}

\smallskip\par\noindent
{\bf Theorem 1.} {\it In every quantum frame  $\Sigma=\Sigma({\cal H})$ the following properties for quantum tests are provable:}

\begin{enumerate}
\item {\it Partial functionality}:
If $s \stackrel{S?}{\to}t$ and $s \stackrel{S?}{\to} v$ then $t = v$.
\item {\it Trivial tests}:
$\stackrel{\emptyset ?}{\to} = \emptyset$ and
$\stackrel{\Sigma?}{\to} = \Delta_{\Sigma},$ where
$\Delta_{\Sigma} = \{(s,s) : s \in \Sigma\}$ is the identity relation on $\Sigma \times \Sigma$.
%\item{ \it{Atomicity}}. States are testable, i.e.
%$\{s\}\in {\cal L}$. \\
%This is equivalent to requiring that ``states can be distinguished by
%tests'', i.e.
%$\mbox{ if } s\not= t \mbox{ then } \exists P\in{\cal L}:\, \, \,   s\perp P, \,  t\not\perp
%P$
\item {\it Adequacy}:
% Testing a true property does not change the state:\\
$\mbox{ if } s \in S \mbox{ then } s \stackrel{S?}{\to} s$
\item {\it Repeatability}:
%Any testable property holds after it has been successfully tested:
{\it If $S \in {\cal L}$ is testable and} $ s \stackrel{S?}{\to} t \mbox{, then } t \in S$
\item {\it Compatibility}:
If $S, T\in {\cal L}$ are testable and $S?;T?$ = $T?;S?$ then $S?;T? = (S\cap T)?$.
\item {\it Self-Adjointness}:
{\it If $s \stackrel{S?}{\to} w \stackrel{T?}{\to} t$ then $t \stackrel{S?}{\to} v \stackrel{W?}{\to} s$, for some $ v \in \Sigma$ and $W \in {\cal L}$. In other words: if $s \stackrel{S?}{\to} w {\to} t$ then $t \stackrel{S?}{\to} v {\to} s$, for some $v \in \Sigma$.}
\item {\it Universal Accessibility}: {\it For all $s,t \in \Sigma$, there exists a state $w \in \Sigma$ such that $s \to w \to t$.
%$\forall s,t \in \Sigma \exists w \in \Sigma \, \, s {\to} w {\to} t$
%\item {\it Unitary Reversibility and Totality}. Basic unitary evolutions are {\it total bijective functions, having as adjoint
%their inverse}:
%$$U; U^{\dagger}=U^{\dagger};U =id$$
%where $id$ is the identity map
%\item {\it Orthogonality Preservation}. Basic unitary evolutions preserve (non) orthogonality:
%Let $s,t,s',t'  \in \Sigma$ be such that $ s \stackrel{U}{\to} s' $and
%$ t \stackrel{U}{\to} t'$.
%Then:\\
%$ s \to t \mbox{ iff } {s' \to t'}$.
}
\end{enumerate}

\par\noindent
{\it Proofs}: $\,$ {\it Partial functionality} follows from the fact that projectors correspond
to partially defined maps in ${\cal H}$.
{\it Trivial tests} follows from the fact that projecting on the empty space yields the empty space and that projecting on the total space doesn't change anything.
{\it Adequacy} follows from the fact that for every $x \in W$ we have that $P_W(x) = x$.
{\it Repeatability} follows from the fact that $P_W(x) \in W$ for every $x \in {\cal H}$.
{\it Compatibility} follows from the fact that if two projectors commute, i.e. $P_W \circ P_V = P_V \circ P_W$, then $P_W \circ P_V = P_{W \cap V}$.
{\it Self-Adjointness} follows from the more general
Adjointness theorem stated below, together with the fact $S?^{\dagger}=S?$.
{\it Universal Accessibility} can be proved by cases: If $s \not \perp t$, i.e. let $s \to t$, then $w = s \Rightarrow s \to s \to t$.  If $s \perp t$, i.e. let $s \not \to t$ then let
$s = \overline{x}, t = \overline{y}$ with $ x, y \in {\cal H}$.  Take the superposition $x + y \in {\cal H}$ of $x $ and $ y $ and note that  $x + y \not = 0$ (since from $x + y = 0 \Rightarrow x = - y \Rightarrow s = t$ which contradicts $s \not \perp t$).  Next observe that $x \not \perp( x + y)$ (Indeed, suppose $x  \perp  (x + y)$ then
$\langle x \mid  x + y \rangle = 0$ and then $\langle x \mid x \rangle + \langle x \mid y \rangle = 0$; but $x \perp y$ implies $\langle x \mid x \rangle = 0$. So from $\langle x \mid x \rangle = 0$ follows that $x = 0$, which yields a contradiction). Similarly, we get $y \not \perp (x + y)$. Taking now $w = \overline{x + y}$, we can see that $w \in \Sigma, s \to w$ and $w \to t$.
%The last two conditions are immediate consequences of the definition of a unitary operator.

%%%%%%%%%%%%%

\smallskip\par\noindent
{\bf Theorem 2.} {\it In every quantum frame $\Sigma({\cal H})$ the following properties for unitary transformation (stated for all $U,U^{\dagger} in {\cal U}$) are provable:}
\begin{enumerate}
\item {\it Functionality: For every state $s \in \Sigma$ we have $\exists!t: s \stackrel{U}{\to} t$}
\item {\it Inverse-adjoint (bijectivity): $s \stackrel{U}{\to} t \stackrel{U^{\dagger}}{\to} w$ implies $s =w$. Similarly,  $s \stackrel{U^{\dagger}}{\to} t \stackrel{U}{\to} w$ implies $s =w$
    }
\end{enumerate}

\smallskip\par\noindent
{\it Proofs: Functionality} follows from the fact that unitary transformations are well-defined on all states, i.e. the kernel of the linear map encoding the transformation is $\emptyset$. {\it Inverse-adjoint} follows from the fact that unitary operators on a Hilbert space have the property that $U^{\dagger} = U^{\mbox{-1}}$.

\smallskip\par\noindent
{\bf Theorem 3.} {\it (Adjointness) $\, \,$ Let $F$ be a linear transformation and let $s,w,t \in \Sigma$ be
states: $\, \, \, \mbox{ If }s \stackrel{F}{\to} w {\to} t$ then there
exists some state $v \in \Sigma$ such that
$t \stackrel{F^{\dagger}} {\to} v {\to} s$.
}
\par\noindent
%{\centerline{
%$$
%\xymatrix{ *{\bullet}  \ar[rr]^{F} & & *{\bullet} \ar @{->}[dd]\\
%& & \\
%*{\bullet} \ar @{..>}[uu]  & & *{\bullet} \ar @{..>}[ll]^{F^{\dagger}}
%}
%$$
%}}
\begin{figure}[!ht]
\centerline{
  \includegraphics[scale=0.5]{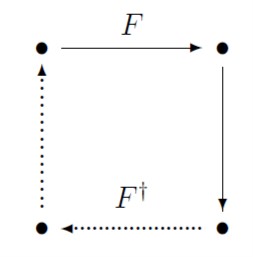}
  }
\end{figure}

\newpage

\smallskip\par\noindent
{\it Proof}:$\,$
To prove this theorem we use the definition of adjointness in a Hilbert space:
$\langle F x \mid y \rangle = \langle x \mid F^{\dagger} y
\rangle$. From this, we get the equivalence:
 $\langle Fx \mid y \rangle = 0$ iff $\langle x, F^{\dagger} y \rangle
= 0$; or, otherwise stated,
$Fx  \perp y $ iff $x  \perp F^{\dagger}y $.
Taking the negation of both sides and using the fact that
the measurement relation $s{\to }t$ is the same as non-orthogonality
$s\not\perp t$, we obtain the equivalence:
$\exists w (\overline{x} \stackrel{F}{\to} \overline{w} \to
\overline{y})$ iff
$\exists v ( \overline{y} \stackrel{F^{\dagger}}{\to} \overline{v} \to
\overline{x})$. This proves the adjointness property. As a
consequence:
\smallskip\par\noindent
{\bf Corrolary 1.} {\it
 For every property $P \subseteq \Sigma$ and every linear map $F$ we have: $P \subseteq [F] \Box \langle F^{\dagger} \rangle \Diamond P$
}

\subsection{Compound System Quantum Frames}
%%%%%%%%%%%%%%%%%%%%%%%%%%%%%%%%%%%%%%%

In this subsection we like to extend the quantum frame presented above
for single systems into a quantum frame for compound systems.  Let $H$
be a Hilbert space of dimension $2$ with basis $\{ \mid 0 \rangle,
\mid 1 \rangle \}$.  We fix a natural number $n\geq 2$ (although later
we will restrict to the case $n\geq 4$), and we put $N=\{1,2,\ldots , n\}$.
A {\it compound-system quantum frame} will be the
quantum frame $\Sigma({\cal H}_n)$ build on a Hilbert space ${\cal H}_n = {
H}^{\otimes n} = {H} \otimes { H} \otimes ... \otimes { H} \mbox{ (n
\, times) }$.

\smallskip\par\noindent
{\bf Notation.}
In fact, we consider all the $n$ copies of $H$ as
distinct (although isomorphic) and denote by $H^{(i)}$ the $i$-th
component of the tensor $H^{\otimes n}$. Also, for any set of indices $I\subseteq N$, we put
${\cal H}_I=H^{\bigotimes I}= \bigotimes_{i\in I}H^{(i)}$. (So, in
particular, ${\cal H}_N={\cal H}_n= {\cal H}$.) We denote by
$\epsilon_i: H\to H^{(i)}$ the canonical isomorphism
between ${\cal H}$ and ${\cal H}^{(i)}$. This notation can be extended
to sets $I\subseteq N$ of indices of length $|I|=k$, by putting
$\epsilon_I: H^{\otimes k}\to {\cal H}_I$ to be the canonical
isomorphism between these two spaces.
For any
vector $\mid x \rangle\in H$, we denote by $\mid x\rangle^{\bigotimes
I}=\bigotimes_{i\in I} \mid x\rangle^{\bigotimes I}$ the corresponding vector
in ${\cal H}_I$ (obtained by tensoring $|I|$ copies of $\mid x\rangle$ ).
Given a set $I\subseteq N$, we say that a state $s\in
\Sigma({\cal H})$ {\it has its $I$-qubits in state} $s'\in
\Sigma({\cal H}_I)$, and write $s_I=s'$, if there exist vectors
$\psi\in s$, $\psi'\in {\cal H}_I$ and $\psi''\in {\cal H}_{N\setminus
I}$ such that $\psi= \mu_I(\psi'\otimes \psi'')$. Note that the state
$s_I$, {\it if it exists, then it is unique} (having the above property).
In particular, when
$I=\{i\}$, we say that state $s$ {\it has as its $i$-th coordinate} the state $s_i\in {\cal H}_{\{i\}}=H^{(i)}$.

We will further denote the vector $\mid 0 \rangle + \mid 1 \rangle$ by
$\mid + \rangle$, and similarly denote
$\mid 0 \rangle - \mid 1 \rangle$ by $\mid - \rangle$.  For the states generated by the vectors in a two dimensional Hilbert space we introduce the following abbreviations:
$+ := \overline{\mid + \rangle}$, $- := \overline{\mid - \rangle}$ , $
0 := \Odiraco$ , $ 1 := \Odirac$.
 In order to refer to the state corresponding to a pair of qubits, we similarly delete the Dirac notation, e.g.
$00 := \overline{\mid 00 \rangle} = \overline{\mid 0 \rangle \otimes \mid 0 \rangle}$.
\par\noindent
The Bell states will be abbreviated as follows:
$\beta_{00} := \overline{\mid 00\rangle +\mid 11\rangle }$ , $\beta_{01} := \overline{\mid 01 \rangle + \mid 10 \rangle }$, $ \beta_{10} := \overline{\mid 00 \rangle -
\mid 11 \rangle}$ , $\beta_{11} = \overline{\mid 01 \rangle - \mid 10 \rangle}$
and \par\noindent
$\gamma := \overline{\mid 00\rangle + \mid 01\rangle + \mid 11\rangle +\mid 10\rangle}$.

The following two results are well-known:

\smallskip\par\noindent
{\bf Proposition 3.}$\,$ {\it
 Let ${H}^{(i)} $ and $ {H}^{(j)}$ be two Hilbert spaces.  There exists a bijective correspondence $\psi$ between the linear maps $F : { H}^{(i)} \to {H}^{(j)}$ and the states of $H^{(i)} \otimes H^{(j)}$.  Given the bases $\{\epsilon_{\alpha}^{(i)}\}_{\alpha}$ and $\{\epsilon_{\beta}^{(j)}\}_{\beta}$  of these spaces, the correspondence $\psi$ is given by the mapping
$ F = \Sigma_{\alpha \beta} \, m_{\alpha\beta} \,  \langle \epsilon_{\alpha}^{(i)} \mid - \rangle . \epsilon_{\beta}^{(j)}$ into the state $\psi(F) = \Sigma_{\alpha \beta} \, m_{\alpha \beta} \, . \epsilon_{\alpha}^{(i)} \otimes \epsilon_{\beta}^{(j)}$.
}

\smallskip\par\noindent
{\bf Proposition 4.} $\,$ {\it Let ${\cal H} = H^{\otimes n}$ and let $W = \{x \, \otimes \mid 0
\rangle^{\otimes (n\mbox{-1})} : x \in H\}$ be given. Any linear map $F:
{\cal H} \to {\cal H}$ induces a linear map ${F_{(1)}: H \to H}$ in a
canonical manner: it is defined as the unique map on $H$ satisfying
$F_{(1)}(x) =$ ${P_W \circ F(x \, \otimes \mid 0 \rangle^{\otimes (n\mbox{-1})})}$.
Conversely, any linear map $G: H \to H$ can be represented as $G = F_{(1)}$ for some linear map $F: {\cal H} \to {\cal H}$.
}

\smallskip
\par\noindent
{\bf Notation.} The above results allow us to specify a compound
state in $H^{(i)} \otimes H^{(j)}$ via some linear map $F$ on ${\cal
H}$. Indeed, if  $F : {\cal H} \to {\cal H}$ is any such linear map,
let $F_{(1)}: H\to H$ be the map in the above proposition; this induces a
corresponding map $F_{(1)}^{(ij)}: H^{(i)} \to H^{(j)}$, by
putting $F_{(1)}^{(ij)} := \epsilon_j \circ F_{(1)} \circ \epsilon_i^{\mbox{-1}}$,
where  $\epsilon_i$ is the canonical isomorphism introduced above (between
$H$ and the $i$-th component $H^{(i)}$ of  $H^{\otimes n}$ ).
Then we denote by $\overline{F}_{(ij)}$ the state
$$\overline{F}_{(ij)} :=
\overline{\psi(F_{(1)}^{(ij)})}$$
given by the above mentioned bijective
correspondence $\psi$ between $H^{(i)} \to H^{(j)}$ and $H^{(i)}
\otimes H^{(j)}$. The following result is also known from the
literature:

\smallskip\par\noindent
{\bf Proposition 5.} $\,$ {\it Let  $F : {\cal H} \to {\cal H}$ be a linear map. Then the state
$\overline{F}_{(ij)}$ is ``entangled according to $F$'';
i.e. if $F_{(1)}(\mid x\rangle)=\mid y\rangle$ and if the state of a
2-qubit system is $\overline{F}_{(ij)}\in H^{(i)}\otimes H^{(j)}$, then any measurement
of qubit $i$ resulting in a state $x_i$ collapses the qubit $j$ to
state $y_j$.
}

\smallskip
\par\noindent
{\bf Notation.}
The notation $\overline{F}_{(ij)}$ can be further extended to
define a property (set of states) $\overline{F}_{ij}\subseteq \Sigma= \Sigma({\cal H})$, by defining it as {\it the set of all
states having the $\{i,j\}$-qubits in the state
$\overline{F}_{(ij)}$ }:
\begin{eqnarray*}
\overline{F}_{ij} \, & = & \, \{s\in \Sigma:
s_{\{i,j\}}=\overline{F}_{(ij)}\} \\
& = & \, \{ \overline{\mu_{\{i,j\}}(\psi\otimes\psi')} :
\psi\in \overline{F}_{(ij)},
\psi' \in
{\cal H}_{N\setminus \{i, j\}} \}\subseteq \Sigma
\end{eqnarray*}

\noindent where $\mu_{\{i,j\}}$ is as above the canonical isomorphism between
${\cal H}_{\{i,j\}}\otimes {\cal H}_{N\setminus \{i,j\}}$. In other
words,
$\overline{F}_{ij}$ is simply the property of an $n$-qubit compound
state of having its $i$-th and $j$-th qubits (separated from the
others, and) in a state that is
``entangled according to $F_{(1)}$''.

\smallskip
\par\noindent
{\bf Local properties}. Given a set $I\subseteq N$, a property
$S\subseteq \Sigma$
is {\it local in $I$ } if it corresponds to a property of the
subsystem
formed by the qubits in $I$; in other words, if there exists some
property $S'\subseteq \Sigma({\cal H}_I)$ such that:
$$
S' =\{s\in \Sigma: s_I \in S'\} $$
or, more explicitly:
$S'=\{\overline{\mu_I(\psi\otimes \psi')}: \overline{\psi}\in S', \psi'\in
{\cal H}_{N\setminus I}\}$.
An {\it example} is the property $\overline{F}_{ij}$, which is
$\{i,j\}$-local.
The family of local properties is closed under union, intersection but {\it not under complementation}.

\smallskip
\par\noindent
{\bf Local transformations}. Given $I\subseteq N$, a linear map $F: {\cal H}\to {\cal
H}$
is {$I$-local} if it ``affects only the qubits in $I$''; in other
words, if there exists a map $G:{\cal H}_I\to {\cal H}_I$ such that:
$$F\circ\mu_I \, (\psi\otimes\psi') \, =\, \mu_I \,( G(\psi) \otimes \psi')$$
A map $F:\Sigma\to \Sigma$ is {\it $I$-local} if it is the map induced
on $\Sigma$ by an $I$-local linear map on ${\cal H}$. {\it Examples} are:
all the tests $S_I ?$ of $I$-local properties; logic gates that affect
only the qubits in $I$, i.e. (maps on $\Sigma$ induced by) unitary
transformations $U_I: {\cal H}\to {\cal H}$ such that for all $\psi,
\psi'\in {\cal H}_I$, we have $U_I\circ \mu_I(\psi\otimes \psi')=
\mu_I ( U(\psi)\otimes \psi')$, for some $U:{\cal H}_I\to {\cal H}_I$. The
family
of local maps is closed under composition.

\smallskip
\par\noindent
{\bf Lemma 2.} {\it The main lemma in \cite{Coecke2} states (in our notation) that, given
a quadruple of distinct indices $i, j, k, l$,  let
$F, G, H, U, V: H\to H$ be single-qubit linear
maps, then we have:
$$\overline{G_{jk}}? \circ V_k \circ U_j\, ( \overline{F}_{ij}\cap
\overline{H}_{kl})\, \subseteq \, \overline{(H\circ U^{\dagger}\circ G \circ
V \circ F)}_{il}$$
}
Using the formalism of {\it entanglement specification networks} introduced in \cite{Coecke2}, this can be encoded in the following diagrammatic representation:

%\centerline{\small{
%$$
%\xymatrix@H=2cm@W=5cm{    *=0{} \ar@{--} [d]   & *=0{}  \ar@{..} [d] &  *=0{} \ar@{.. } [d] %&*=0{} \ar@{--} [d]\\
%*=0{\bullet} \ar@/^/ @{-} [rrr]^{H o V^\dagger o G o U o F} \ar@{--} [d] \ar@{.} [r] & *=0{}  %\ar@{.} [r] \ar@{..} [d] &  *=0{}  \ar@{.} [r] \ar@{..} [d] & *=0{\bullet}  \ar@{--} [d]  \\
%*=0{} \ar@{--} [d] \ar@{.} [r] & *=0{\bullet}  \ar@{-} [r]^{G} \ar@{--} [d] &  *=0{\bullet}  %\ar@{.} [r] \ar@{--} [d] & *=0{}  \ar@{--} [d]  \\
%*=0{} \ar@{--} [d] \ar@{.} [r] & *++[oo][F-:<0pt>]{U}  \ar@{.} [r] \ar@{--} [d] &
% *++[oo][F-:<0pt>]{V}  \ar@{.} [r] \ar@{--} [d] & *=0{}  \ar@{--} [d]  \\
% *=0{\bullet} \ar@{-} [r]^{F}  \ar@{--} [d] &*=0{\bullet}  \ar@{.} [r] \ar@{--} [d] &  %*=0{\bullet} \ar@{-} [r]^{H}  \ar@{--} [d] &*=0{\bullet} \ar@{--} [d] \\
%  *=0{}   & *=0{}   &  *=0{} &*=0{} }
%$$
%}}

\begin{figure}[h]
\centerline{
  \includegraphics[scale=0.5]{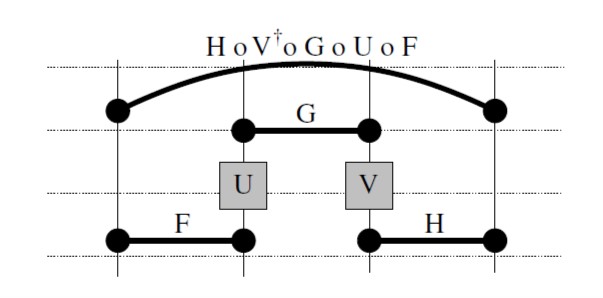}
  }
\end{figure}
\par\noindent
\cite{Coecke2} and \cite{AbramskiCoecke} use this as the main tool in explaining teleportation,
quantum gate teleportation
and many other quantum protocols. We will use this work in our logical
treatment of such protocols, by taking this lemma as one of our main axioms.

Observe that in the above Lemma, the order in which the operations
$U_j$ and $V_k$ are applied is in fact {\it irrelevant}. This is a consequence of
the following important property of local transformations:

\smallskip
\par\noindent
{\bf Proposition 6.} {\it (Compatibility of local transformations affecting different
sets of qubits)}
\par\noindent
{\it If $I\cap J=\emptyset$, $F_I$ is an $I$-local map and
$G_J$ is a $J$-local map, then we have:
$$F_I \circ G_J = G_J \circ F_I$$
}

Another important property of local maps (on {\it states}) is:

\smallskip\par\noindent
{\bf Proposition 7.} {\it ( ``Agreement Property'')} {\it Let $F_I, G_I: \Sigma\to \Sigma$
be two $I$-local maps
on states, having the same domain\footnote{The domain of a map is
defined by $dom(F)=\{s\in \Sigma: F(s) \mbox{ is defined } \}$. If $F'$
is the corresponding linear map on ${\cal H}$, this means that
$dom(F)=\{\overline{\psi}: F'(\psi)\not=0\}$.}:
$dom(F)=dom(G)$. Then their output-states agree on all non-$I$ qubits,
i.e.:
$$F(s)_{J} = G(s)_{J}$$
for all $s\in \Sigma$ and all $J$ such that $I \cap J = \emptyset$. (We take this equality to imply in particular that the right-hand is defined iff the left-hand is also defined.)
}
%%%%%%%%%%%%%%%%%%%%%%%%%%%
\smallskip
\par\noindent{\bf Dynamic Characterizations of Main Unitary Transformations.}
%%%%%%%%%%%%%%%%%%%%%%%%%%%%%%%%%%%%%%%
\smallskip\par\noindent
It is well-known that a linear operator on a vector space in a given
Hilbert space is {\it uniquely determined} by the values it takes on
the vectors of an (orthonormal) basis.  An important observation is
that this fact is no longer ``literally true'' when we move to
``states'' as one-dimensional subspaces instead of vectors. The reason
is that ``phase''-aspects (or, in particular, the signs ``$+$'' and
``$-$'') are not ``state'' properties in our setting.  In other words,
two vectors that differ only in phase, i.e $x = \lambda y$ where $\lambda$ is a complex number
with $\mid \lambda \mid = 1$, belong to the same subspaces, so they correspond to the
same state $\overline{x} = \overline{y}$.
\medskip\par\noindent
{\bf Example 1.}$\,$ {\it (Counterexample)}
 $\, \,$
Consider a 2 dimensional Hilbert space in which we denote the basis
vectors by $\mid 0 \rangle$
and $\mid 1 \rangle$, a transformation $I$ is given by $I(\alpha {\mid
0 \rangle} + \beta {\mid 1 \rangle}) = \alpha \mid 0
\rangle + \beta \mid 1 \rangle$; and a transformation $J$ is given by $J(\alpha {\mid 0 \rangle} + \beta {\mid 1 \rangle}) = \alpha \mid 0 \rangle - \beta \mid 1 \rangle$.  Although $I $ and $J$ induce different operators on states, these operators map the basis states to the same images:
\par\noindent
$I(0) = \overline{I(\mid 0 \rangle)} = 0 = \overline{J(\mid 0 \rangle)} = J(0)$,
$I(1) = \overline{I(\dirac)} = 1 = \overline{- \dirac} = \overline{J(\dirac)} = J(1)$.
But of course we do distinguish the subspaces generated by different superpositions:
$I(+) = \overline{\diraco + \dirac}  = + \not = - = \overline{\diraco - \dirac} = J(+)$.

\smallskip\par\noindent
{\bf Proposition 8.} $\,$ {\it
A linear operator on the state space $\Sigma({\cal H}_1)$ of a 2 dimensional Hilbert space is
uniquely determined by its images on the states: $\Odiraco, \Odirac, \overline{\mid + \rangle}$.
}

\smallskip\par\noindent
{\bf Corollary 2.} $\,$ {\it A linear operator on the state space $\Sigma({\cal H}_n)$ of the space ${\cal H}_n$ is uniquely determined by its
images on the states:
$$\{\overline{ \mid x \rangle_1 \otimes ... \otimes \mid x \rangle_n} : \mid x \rangle_i \in \{ \mid 1\rangle_i, \mid 0\rangle_i, \mid +\rangle_i \}\}$$
}

\smallskip
\par\noindent
In the definition of a quantum frame given above, we introduced the set ${\cal U}$
as the set of unitary transformations for single systems.
For compound systems the set ${\cal U}$ will be extended with the kind of operators
that are active on compound systems.  Following the quantum computation literature, we take
${\cal U} = \{ X, Z, H, CNOT, ...\}$ where $X, Z$ and $H$ are defined by the following table:
\bigskip\par\noindent
\centerline{\begin{tabular}{r||r|r|r}
 & 0 & 1 & + \\
 \hline \hline
 X& 1 & 0 & + \\
 \hline
 Z & 0 & 1 & - \\
 \hline
 H & + & - & 0
 \end{tabular}}
\bigskip\par\noindent
The transformation $CNOT$ is given by the table:
\bigskip\par\noindent
\begin{tabular}{r||r|r|r|r|r|r|r|r|r}
&        $00$&  $01$&  $0+$& $11$& $10$& $1+$ &$+0$         &$ +1$        &$++ $\\
\hline \hline
$CNOT $& $00$& $01$&   $0+$& $10$& $11$& $1+$ &$\beta_{00}$ &$\beta_{01}$ &$ \gamma $
 \end{tabular}

%%%%%%%%%%%%%%%%%%%%%%%%%%%%%%%%%%%%%%%
\section{Syntax of $LQP$}
%%%%%%%%%%%%%%%%%%%%%%%%%%%%%%%%%%%%%%%

{\bf The Basic Language of $LQP$}
\smallskip\par\noindent
To build up the language of $LQP$, we are given a natural number $n$,
and we put $N=\{1, 2, \ldots, n\}$. We
start from a set ${\cal Q}$ of {\it propositional variables}, together with an {\it arity map},i.e. every $p \in {\cal Q}$ has an arity $k\leq n$; a
 set ${\cal C}=\{+,1,...\}$
of {\it propositional
constants}; and a set ${\cal U}=\{CNOT_2,X_1,H_1,Z_1,...\}$ of constants, denoting {\it basic programs}, to be interpreted as {\it unitary transformations}; each such program comes also with an arity $k \leq n$. The syntax of $LQP$ is an extension of the classical syntax for $PDL$, with a set of propositional {\it formulas} and a set of {\it programs}, defined by mutual induction:
{\small{
\beqa
\begin{array}{lllllllllllllll}
\varphi & ::=   &p_I  & \mid & c_i & \mid & \overline{\pi}_{i,j} & \mid & \neg \varphi  &  \mid & \varphi \wedge \varphi & \mid & [\pi] \varphi &  \\
\pi & ::=        & \top & \mid & \varphi? &  \mid & U_I & \mid & \pi^{\dagger} & \mid  & \pi \cup \pi  & \mid & \pi ; \pi & \mid & \pi^{*}
\end{array}
\eeqa
}}

\par\noindent
Here, we take $I$ to denote sequences of distinct indices in $N=\{1,2,\ldots, n\}$.
In the above syntax, $p_I$ and $U_I$ are well-formed terms iff the arity $k$ of $p$, or of $U$, matches the length of the sequence, i.e. $k=|I|$. In the semantics we will interpret $p$ to be a physical property of the system of $|I|$ qubits, and the sentence $p_I$ as saying that the qubits with indices in $I$ have the property $p$ consisting of $k=|I|$ relevant basic states which are specifically the ones labeled corresponding to the numbers in the subset $I$. Similarly, in the semantics it will become clear that every member of ${\cal U}$ encodes a specific quantum logical gate and the subscript $I$ in $U_I$ will then indicate on which qubits the gate is active. When the arity of a variable $p$ is $n$, then we skip the subscript, and simply write $p$ instead of $p_n$.
%%%%%%%%%%%%%%%%

For a given propositional constant $c \in {\cal C}$, we interpret the sentence $c_i$ as saying that
``the i-th-qubit is in the state $|c>$''. Note that $1$ as a logical (characterizing the qubit $|1>$) is
different from the propositional formula $\top$ (verum) which we formally introduce later in
this section, to denote the ``top'' element of the lattice of properties. This, in its turn, is also
different from the {\it program} $\top$, introduced in the syntax above, which will simply denote
the trivial program, relating any two states.
%%%%%%%%%%

\smallskip
\par\noindent
{\bf Extending the Basic Language of $LQP$}.
\smallskip\par\noindent
We extend our language by defining the operations for a {\it classical disjunction} and a {\it classical implication}  in the usual way, i.e.
$\varphi \vee \psi := \neg (\neg \varphi \wedge \neg \psi)$,
$\varphi \to \psi := \neg \varphi \vee \psi$. We introduce constants
{\it verum}
$\top := 1_1 \vee \neg 1_1$, and {\it falsum}
$\bot := 1_1 \wedge \neg 1_1$.
We define the {\it classical dual} of $[\pi] \varphi$ in the usual way as
$\langle \pi \rangle \varphi := \neg [\pi] \neg \varphi$ ;
the {\it measurement modalities} $\Box$ and $\Diamond$ that are known in the quantum logic literature can be defined in $LQP$ by putting
$\Diamond \varphi : = \langle \varphi ? \rangle \top$ and $\Box \varphi
:= \neg \Diamond \neg \varphi$.
The {\it orthocomplement} is defined as $\sim \varphi := \Box \neg \varphi$, or equivalently as $\sim \varphi := [\varphi?]\bot$.
By means of the orthocomplement we define new propositional constants $0_i := \sim 1_i$ and $-_i := \sim +_i$, and a binary operation for
 {\it quantum join}
$\varphi \sqcup \psi := \sim (\sim \varphi \wedge \sim \psi)$.
This expresses {\it superpositions}: $\varphi\sqcup \psi$ is true at
any state which is a superposition of states satisfying
$\varphi$ or $\psi$. We can also define the {\it quantum dual} of a modality $[\pi]\psi$ as $<\pi^{\sim}>\psi :=\sim [\pi]\sim \psi$. Finally we put $<\pi>^{\mbox{-1}}\psi := <(\pi^{\dagger})^{\sim}>\psi$. As we'll see, this captures the {\it strongest post-condition} ensured by applying program $\pi$ on a state satisfying (a precondition) $\psi$.

\smallskip\par\noindent
{\bf Testable formulas.} We call a program $\pi$ {\it deterministic} if $\pi$ is constructed
without the use of non-deterministic choice $\cup$ or iteration $*$. Next we define the set of {\it testable formulas} $\varphi_t$ of $LQP$ to be a subset of the above given language, constructed by induction in the following way:

{\small{
\beqa
\begin{array}{llllllllllll}
\varphi_t & ::=   & \bot  & \mid & c_i & \mid & \overline{\pi}_{i,j} & \mid &  \varphi_t \wedge \varphi_t  &  \mid & [\pi] \varphi_t
\end{array}
\eeqa
}}
\par\noindent
where $\pi$ is any {\it deterministic program}. Observe that the construction of $\pi$ might involve non-testable formulas. In particular, for an arbitrary (not necessarily testable) formula $\varphi$, remark that $[\varphi?]\psi_t$ is a testable formula.

\smallskip\par\noindent
{\bf Proposition 9.} {\it For any formula $\varphi$ in $LQP$, $\sim \varphi$ and $\Box \varphi$ are testable formulas.}

\smallskip
\par\noindent
\textbf{Local formulas and local programs}. We would like to isolate {\it local} formulas and programs, i.e. the ones that ``affect only
the qubits in a given set $I \subseteq N$''. These formulas will express local properties (in the sense defined above). When we want to stress that a formula or program is local, we denote them with $\varphi_I$ or $\pi_I$. The definition is:

{\small{
\beqa
\begin{array}{lllllllllllll}
\varphi_I & ::=   & p_J         & \mid & c_i & \mid & \overline{\pi}_{i,j} & \mid &    \varphi_I \vee \varphi_I  &  \mid  & \varphi_I \wedge \neg \varphi_I & \mid & \varphi_I \wedge [\pi_I] \varphi_I\\
\pi_I      & ::=  & \varphi_I ? & \mid & U_J & \mid &  \pi_I;\pi_J          & \mid &
\pi_I \cup \pi_I          & \mid  &   \pi_I^{*}
\end{array}
\eeqa
}}
\par\noindent
with $i,j \in I, J \subseteq I$. Observe that local formulas are not closed under negation: this is because the complement of a local property is not necessarily a local property. But instead
they are closed under set-theoretic difference, disjunction, and also conjunction: this is
because $\varphi \wedge \psi$ is equivalent to $\varphi \wedge \neg (\varphi \wedge \neg \psi)$.

\smallskip\par\noindent
{\bf Relabeling local formulas and programs}. When we label a local formula $\varphi_I$ or a
local program $\pi_I$ with a sequence of indices $I$, we can of course take any other sequence
$J$ of indices, with $|J| = |I|$, and substitute all the $I$ indices in our formula (program) with
the corresponding $J$ indices; we denote by $\varphi_J$ , and respectively $\pi_J$ , the corresponding
formula, or program.

\smallskip\par\noindent
{\bf Notation.} {\it The unary map induced by a program:} We want to capture in our syntax the
construction $F_{(1)}$, by which a linear map $F$ on $H^{\otimes
n}$ was used to describe a unary map
$F_{(1)}$ on $H$. For this, we put: $0_i!$ := $0_i? \cup (1_i?;X_i)$, and $0_I!$ := $0_{i_1}!; 0_{i_2}!; ... ; 0_{i_k}!$,
where $I = (i_1, i_2, ... , i_k)$. This maps any qubit in $I$ to $0$. Similarly, we put; $0_I? :=
(0_{i_1} \wedge 0_{i_2} \wedge ... \wedge 0_{i_k} )?$. Finally we define:
%%%%%%%%%%%%
$$\pi_{(i)} :+ 0_{N\setminus \{i\}}!;\pi;0_{N\setminus \{i\}}?$$
\par\noindent
This is the map we need (which encodes a single qubit transformation). In fact, we shall
only use $\pi_{(1)}$ in the rest of this paper.

%%%%%%%%%%%%
\section{Semantics of $LQP$}
%%%%%%%%%%%%%%%%%%%%%%%%%%%%%%%%%%%%%%%

An {\it $LQP$-model} is a {\it quantum frame equipped with a valuation function}, mapping each propositional variable $p$ with arity $k$ into a set
$\M p \M \subseteq \Sigma(H^{\otimes k})$ of $k$-qubit states. Give na sequence $I$ of length $i$ of indices, let $\epsilon$ be the canonical isomorphism between $H^{\otimes k}$ and $H^{\otimes I}$.
\par\noindent
We will use the valuation map to give an interpretation $\M
\varphi\M \, \subseteq \Sigma$
to all our formulas, in terms of properties of our $n$ qubit system,
i.e. sets of states in $\Sigma = \Sigma({\cal H})$. In the same time, we
give an interpretation $\M \pi \M \, \subseteq  \Sigma \times\Sigma$ to all our programs,
in terms of binary relations between states. The two interpretations are defined by {\it mutual recursion}.
\smallskip\par\noindent
{\bf Interpretation of Programs:} The basic programs $U_I$ , with $|I| = k$, come from a
list of corresponding $k$-bit unitary transformations $U : H^{\otimes k} \to
H^{\otimes k}$. We take $\M U_I \M$ to
be the (map on states induced by the) unique linear map on ${\cal H}$ such that:

$$\M U_I \M \circ \mu_I (\psi \otimes \psi') := \mu_I(\epsilon_I \circ U \circ \epsilon_{i}^{\mbox{-1}}(\psi) \otimes \psi')$$

\par\noindent
for every $\psi \in {\cal H}_I, \psi'\in {\cal H}_{N\setminus I}$. Here, recall that $\epsilon_I$ is the canonical isomorphism between $H^{\otimes k}$ and ${\cal H}_I$, and $\mu_I$ i the canonical isomorphism between ${\cal H}_I \otimes {\cal H}_{N\setminus I}$ and ${\cal H}$.
\par\noindent
As for the others:

\begin{eqnarray*}
\begin{array}{lllllll}
\M \top \M \, & := & \, \Sigma\times\Sigma& , &
\M \varphi ?\M \, & := & \, \M \varphi \M ? \\
\M \pi_1 \cup \pi_2 \M \, & := & \, \M \pi_1 \M \cup \M \pi_2 \M  & , &
\M \pi^{*} \M  \, & := & \, \M\pi\M ^{*} \\
\M \pi_1 ; \pi_2 \M  \, & := & \, \M \pi_1 \M \circ \M \pi_2 \M & , &
\M U^{\dagger}_{I} \M \, & := & \,  \M U_I \M^{\mbox{-1}} \\
\M (\pi^{\dagger})^{\dagger} \M \, & := & \, \M \pi \M & , & \M (\pi_1;\pi_2)^{\dagger} \M \, & := & \, \M \pi_2^{\dagger};\pi_1^{\dagger} \M \\
\M (\pi \cup \pi_2)^{\dagger} \M \, & := & \, \M (\pi_1)^{\dagger} \cup (\pi_2)^{\dagger} \M \, &, & \, \M (\pi^{*})^{\dagger} \M \, & := & \, \M (\pi^{\dagger})^{*} \M
\end{array}
\end{eqnarray*}
\par\noindent
where $R^{*}$ is the reflexive-transitive closure of relation $R$. Note that {\it deterministic programs}
have as interpretations $|| \pi ||$ (maps on states which are induced by) {\it linear maps} on ${\cal H}$.
The interpretation $|| \pi ||$ allows us to extend the notation $\stackrel{\pi}{\to}$
to all programs, by putting:
\par\noindent
$s \stackrel{\pi}{\to} t$ iff $(s, t) \in || \pi ||$.

\smallskip\par\noindent
{\bf Interpretation of Formulas.} We give the interpretation here first for all except
propositional variables $p_i$ and entangled state formulas $\overline{\pi}_{ij}$ :
\begin{eqnarray*}
\begin{array} {lllllll}
\M \varphi \wedge \psi \M  \, & = & \,  \M \varphi \M \,  \cap \, \M \psi \M & , &
\M \neg \varphi \M  \, & = & \, \Sigma \backslash \M \varphi \M \\
\M 1_i \M  \, & = & \,  1_i & , &
\M + \M  \, & = & \, +_i
\end{array}
\end{eqnarray*}

\par\noindent
and finally $\M [\pi]\varphi\M$ =$\{s \in \Sigma | \forall t: s \stackrel{\pi}{\to} t \Rightarrow t \in \M \varphi \M\}$.
\par\noindent
The last clause obviously defines {\it the weakest precondition} $[\pi]\varphi$ ensuring that (postcondition) $\varphi$ will be satisfied after executing program $\pi$. As for the propositional variables, we
put:
$$ \M p_I \M \, = \, \{s \in {\cal H}: s_I \in \epsilon_I(\M p \M)\}$$
$$ \quad \quad \quad \quad \quad \quad \quad \quad \quad \quad  = \, \{\overline{\mu_I(\epsilon_I(\psi) \otimes \psi')} : \overline{\psi} \in \M p \M, \psi' \in {\cal H}_{N \setminus I}\}$$

\par\noindent
where $\epsilon_I$ and $\mu_I$ are the above-mentioned canonical isomorphisms, and $s_I$ is (as defined
above) the state of the qubits in $I$. So the meaning of $p_I$ is that the system of qubits with
indices in $I$ is separated from (i.e. non-entangled with) the rest of the system, and that
moreover this system has the property expressed by $p$.

The interpretation of $\overline{\pi}_{ij}$ , for {\it deterministic programs} $\pi$, is given by the construction
$\overline{F}_{ij}$ above. Since the interpretation $|| \pi ||$ of a deterministic program is a linear map on ${\cal H}$,
we know, by the results mentioned above, that the map $F_{(1)}$ can be used to specify a set of
compound states $\overline{F}_{ij} \subseteq H$. This is our intended interpretation for $\overline{\pi}_{ij}$:

$$\M \overline{\pi}_{ij} \M := \overline{ \M \pi \M }_{ij} $$

\par\noindent
For the program $\top$, we put: $|| \overline{\top} ||:= \{s \in \Sigma : s_{i,j} \mbox{ is defined} \} = \{\overline{\mu_{\{i,j\}}(\psi \otimes \psi')} :   \psi \in
{\cal H}_{\{i j\}}, \psi \in {\cal H}_{N\setminus\{i,j\}}\}$, i.e. the property of having the ${i, j}$-qubits in a separated state from the others. This can be extended to other programs in the natural way, by putting e.g.
\par\noindent
$\M \overline{\pi \cup \pi'_{ij} } \M := \M \overline{\pi}_{ij} \cup \overline{\pi '}_{ij} \M$ etc.

\smallskip\par\noindent
{\bf Proposition 10.} {\it The interpretation of any testable formula is a testable property. The
interpretation of an $I$-local formula (or deterministic program) is
an $I$-local
formula (or $I$-local linear map on states).}

\smallskip\par\noindent
{\bf Lemma 3.} $\M\sim \varphi\M=\M\varphi\M^{\perp}$,
$\M[\varphi?]\psi\M=[\M\varphi\M ?]\M\psi\M$, $\M\Box\varphi\M=\Box\M\varphi\M$,
$\overline{\M \varphi\M }=\M \sim\sim \varphi\M$

\smallskip\par\noindent
{\bf Proposition 11.} {\it The following are equivalent, for every formula $\varphi$:} \\
\begin{tabular}{ll}
$1.$ & $ \,  \M\varphi\M$  is testable \\
$2. $ & $\, \varphi$  is semantically equivalent to $\Box\Diamond\varphi$\\
$3. $ & $\, \varphi$  is semantically equivalent to some formula $\Box\psi$\\
$4. $ & $\, \varphi$  is equivalent to some formula $\sim \psi$
\end{tabular}

%%%%%%%%%%%%%%%%%%%%%%%%%%%%%%%%%%%%%%%
\section{Axioms for $LQP$}
%%%%%%%%%%%%%%%%%%%%%%%%%%%%%%%%%%%%%%%

First, we  admit {\it all the axioms and rules} of {\bf classical
$PDL$}, except for the ones concerning tests $\varphi?$ In particular,
we have the basic axiom and rule for sentence involving {\it modalities} $[\pi]$, stated for elementary sentences and basic programs:

\smallskip\par\noindent
{\bf Kripke Axiom. } $\, \,  \vdash \, [\pi](p \to q) \rightarrow ([\pi] p \rightarrow [\pi] q)$
\par\noindent
{\bf Necessitation Rule. } $\, \, \mbox{ if }  \vdash p \mbox{ then } \vdash [\pi] p$
\smallskip\par\noindent
Considering $\Box p$, we introduce the following axioms:
\par\noindent
{\bf Test Generalization Rule. }
$\, \, \mbox{ if}  p \rightarrow [q?]r  \mbox{ for all } q, \mbox{ then} \vdash \, p \to \Box r$
\par\noindent
{\bf Testability Axiom. }  $\, \, \vdash  \Box p \rightarrow [q?]p$

\smallskip\par\noindent
Testability can be stated in its dual form by means of $\langle q?
\rangle p \rightarrow \Diamond p$ or
equivalently as $\langle q? \rangle p \rightarrow \langle p? \rangle
\top$.
This dual formulation of Testability allows us to give a
straightforward interpretation:
if the property associated to $p$ can be actualized by a measurement
(yielding an output state satisfying
 $p$), then we can directly test the property $p$ (by doing a measurement for $p$). The Test Generalization Rule encodes the fact that $\Box$ is a universal quantifier over all possible measurements.

\smallskip\par\noindent
Other $LQP$-axioms are:

\smallskip\par\noindent
\begin{tabular}{lll}
{\bf Partial Functionality.}  $\, \,$  &$ \vdash $& $\neg [p?]q \rightarrow  [p?] \neg q $\\
{\bf Adequacy.} &$ \vdash $&$ p \wedge q \rightarrow \langle p? \rangle q $
\\
{\bf Repeatability.}  $\, \,$ &$ \vdash $&$ [\psi_t ?]q \to [p?] \neg q $ for all {\it testable} formulas $\psi_t$
\\
{\bf Universal Accessibility.} $\, \,$ &$ \vdash $&$ \langle \pi \rangle \Box \Box p \rightarrow [\pi'] p $
\\
{\bf Unitary Functionality.} $\, \,$ &$ \vdash $ & $\neg [U]q \leftrightarrow  [U] \neg q $
\\
{\bf Unitary Bijectivity 1.} $\,  \,$ &$ \vdash $&$ p \leftrightarrow [U;U^{\dagger}] p $
\\
{\bf Unitary Bijectivity 2.} $\, \,$ &$ \vdash $&$ p \leftrightarrow [U^{\dagger};U] p $
\\
{\bf Adjointness.} $\, \,$  &$ \vdash $&$ p \rightarrow [\pi] \Box \langle \pi^{\dagger}\rangle \Diamond p $
\end{tabular}
\par\noindent
{\bf Substitution Rule.} $\,\,$ From $\vdash \Theta$ infer $\vdash \Theta[p_I\backslash \varphi_I]$
\par\noindent
{\bf Compatibility Rule.} $\,\,$ For all {\it testable} formulas $\psi, \varphi$ and every variable $p \not \in \varphi, \psi$:
$$\mbox{ From } \vdash <\varphi?;\psi?>p \to <\psi?;\varphi?>p \mbox{ infer } \vdash <\varphi?;\varphi?>p \to <(\varphi \wedge \psi)?>p$$
%\par\noindent
%Before going to other axioms, we note some immediate consequences:
%\bpr $\, \vdash \Box p \rightarrow p, \, \vdash p \rightarrow \Box \Diamond p$
%\epr

\smallskip\par\noindent
{\bf Proposition 12.} {\it (Quantum Logic, Weak Modularity or Quantum Modus Ponens) $\, \,$
All the axioms and rules of traditional Quantum
Logic are satisfied by our testable formulas. In
particular, from our axioms one can prove ``Quantum Modus
Ponens''\footnote{This explains why the weakest precondition
$[\varphi?]\psi$ has been taken as the basic implicational connective
in traditional Quantum Logic, under the name of ``Sasaki hook'',
denoted by $\varphi\stackrel{S}{\to}\psi$.}
$\varphi \wedge [\varphi?]\psi \vdash \psi$. In its turn, this rule
is equivalent  to the condition known in quantum logic as Weak Modularity, stated as follows:
$\varphi \wedge (\sim \varphi \sqcup (\varphi \wedge \psi)) \vdash \psi$.
}

\smallskip\par\noindent
{\bf Theorem 4.} {\it (Soundness, Expressivity, Completeness of the above axioms with respect to $PDL$ frames) In the presence of (axioms of classical logic, plus) Kripke's Axioms, Necessitation, Test Generalization, Testability and Substitution Rule, all the other axioms above are sound and expressive with respect to the corresponding semantic conditions mentioned in Section 2 above. More precisely:
any of these axioms is valid on a $PDL$ frame iff the
corresponding semantic condition is satisfied by the frame. Moreover, the system given by
the above axioms is complete for the class of $PDL$ frames satisfying all the corresponding
semantic conditions.}

\smallskip\par\noindent
{\bf Proposition 13.} {\it The formula $<\pi>^{\mbox{-1}}\varphi$ expresses the strongest testable postcondition ensured by executing program $\pi$ on any state satisfying (precondition) $\varphi$. In other words: for every testable $\psi$ the following are equivalent:}
\begin{enumerate}
  \item $\vdash <\pi>^{\mbox{-1}} \varphi \to \psi$
  \item $\vdash \varphi \to [\pi]\psi$
\end{enumerate}
\par\noindent
{\it Moreover, in the context of the other axioms, this equivalence is itself equivalent to the
Adjointness Axiom.}

\smallskip\par\noindent
{\bf Basic Axioms for constants} $(0,1,+,-)$
\par\noindent
The first axiom says that $c_i$'s are ``states'' in the $i$-th part of the system, i.e. they are atomic
properties, which determine completely whether any other property is jointly satisfied. We
state in a {\it weak}, as well as in stronger version:

\smallskip\par\noindent
{\bf Atomicity} (weak version). For all $c \in \{0,1,+,-\}$: $\, \vdash c_i \wedge p_i \to \Box \Box(c_i \to p_i)$
\par\noindent
{\bf Atomicity} (strong version). For all $c \in \{0,1,+,-\}$:
\par\noindent
$\vdash \bigwedge_{i\in I} c_i \wedge p_I \to \Box \Box (\bigwedge_{i \in I} c_i \to p_I)$
\smallskip\par\noindent
The following axioms state that $+_i$ and $\mbox{-}_i$ are proper superpositions of $0_i$ and $1_i$:
\smallskip\par\noindent
{\bf Proper Superposition Axioms:} $\vdash +_i \to \Diamond 0_i \wedge \Box 1_i$ and
$\vdash \mbox{-}_i \to \Diamond 0_i \wedge \Diamond 1_i$.
\smallskip\par\noindent
Next two axioms assert that $1$ and $+$ are {\it testable} properties:
\smallskip\par\noindent
{\bf Constants are testable.} $\vdash \Box\Diamond 1_i \to 1_i$ and $\vdash \Box\Diamond +_i \to +_i$.
\smallskip\par\noindent
{\bf Determinacy Axiom of Deterministic Programs.} For deterministic programs $\pi,\pi'$:
\par\noindent
$\vdash(\Box\Box\bigwedge_{c^{(1)},...,c^{(n)}\in \{0,1,+\}^{n}} (<\pi>^{\mbox{-1}}(c_1^{(1)} \wedge ... \wedge c_n^{(n)}) \leftrightarrow <\pi>^{\mbox{-1}}(c_1^{(1)}\wedge ...\wedge c_n^{(n)}))) \to (<\pi>p \leftrightarrow<\pi'>p)$
\par\noindent
This expresses the above-mentioned property of linear operators on ${\cal H}$ of being uniquely
determined by their values on all the states $|x>_1 \otimes ... |x>_n$, with $|x>_i \in \{|0>_i, |1>_i, |+>_i\}$.
\smallskip\par\noindent
{\bf Agreement Axiom.} If two $I$-local programs $\pi,\pi'$ have the same domain, then their
output states agree on all non-$I$ qubits: i.e. if $I \cap J = \emptyset$ then
\par\noindent
$\Box\Box(<\pi_I>\top \leftrightarrow <\pi'>\top ) \to (<\pi_I>p_J \leftrightarrow <\pi'_I>p_J)$
\smallskip\par\noindent
{\bf Compatibility of programs affecting different sets of qubits.} If $I \cap J = \emptyset$ then
\par\noindent
$\vdash [\pi_I;\pi_J]p \leftrightarrow [\pi_J;\pi_I]p$
\smallskip\par\noindent
{\bf Entanglement Rule.} From $\vdash p_1 \to [\pi_{(1)}]q_1$ infer $\vdash \overline{\pi_{ij}} \to [p_i?]q_j$
\smallskip\par\noindent
{\bf Entanglement Composition Axiom}. For distinct indices
$i,j,k,l$, programs $\pi, \pi', \pi''$ and local $\{1\}$-programs $\sigma_1, \rho_1$
we have:
\\ $\vdash \overline{\pi}_{ij} \wedge \overline{\pi'}_{kl}\rightarrow
[\sigma_j; \rho_k; \overline{\pi''}_{jk}?]
\overline{(\pi; \sigma_1; \pi''; \rho_1^{\dagger};\pi')}_{il}$
\smallskip\par\noindent
{\bf Trivial Entanglement} $\vdash p_{i,j} \to \overline{\top}_{ij}$ This says that separation of the $i,j$-qubits implies their trivial entanglement.

%%%%%%%%%%%%%%%%%%%

\smallskip\par\noindent
{\bf Theorem 5.} {\it (Teleportation Property)}. {\it If $\varphi_1$ is a $1$-local testable property and if $\vdash \varphi_1 \to [\pi_{(1)};\sigma_{(1)}]q_1$, then $\vdash \varphi_1 \wedge \overline{\sigma}_{23} \to [\overline{\pi}_{12}?]q_3$.}

\smallskip\par\noindent
{\it Proof:} We apply the Entanglement Composition Axiom, taking $i = 4, j = 1, k =
2, l = 3$, and substituting the programs $\top$ for $\pi$, $\sigma$ for $\pi'$, $\pi$ for $\pi''$, $\varphi_1?$ for $\sigma_1$, and
$id_1 = X_1;X_1$ for $\rho_1$. We obtain: $\vdash \overline{\top}_{41} \wedge \overline{\sigma}_{23} \to [\varphi_1?;id_2;\overline{\pi}_{12}?]\overline{(\top;p_1?;\pi;id_1^{\dagger};\sigma)}_{43}$.
On the other hand, we have $\vdash \varphi_1 \wedge \overline{\sigma}_{23} \to [0_4!](p_1 \wedge \overline{\top}_{41} \wedge \overline{\sigma}_{23})$ (since $0_4!$ is $4$-local and has the same domain as $id_4$, so by Agreement Axiom it agrees with $id_4$ on non-$4$ qubits, thus preserving $\varphi_1$ and $\overline{\sigma}_{23}$; but also $\vdash [0_4!]0_4$ and using the Trivial Entanglement Axiom, we get the conclusion). From these two together, we obtain:
$\vdash \varphi_1 \wedge \overline{\sigma}_{23} \to [0_4!][\overline{\pi}_{12}?]\overline{(\top;\varphi_1?;\pi;id_1^{\dagger};\sigma)}_{43}$. But on the other hand, we have $\vdash \overline{(\top;\varphi_1?;\pi;id_1^{\dagger};\sigma)}_{43} \to [0_4?]q_3$. (This is because we assumed $\vdash \varphi_1 \to [\pi_{(1)};\sigma_{(1)}]q_1$, from which it follows that $\vdash 0_1 \to [\top;\varphi_1?;\pi_{(1)};id_1^{\dagger};\sigma_{(1)}]q_1$, using the fact that $id^{\dagger}=id$ and $\vdash [\varphi_1?]\varphi_1$, by Repeatability axiom and the testability of $\varphi_1$. Apply now Entanglement Rule, obtaining the above conclusion.) From these two we get that: $\vdash \varphi_1 \wedge \overline{\sigma}_{23} \to [0_4!;\overline{\pi}_{12}?;0_4?]q_3$. The desired conclusion follows from the Agreement Axiom and the fact that $0_4!;\overline{\pi}_{12}?;0_4?$ and $\overline{\pi}_{12}?$ are $\{1,2,4\}$-local programs with the same domain.

\smallskip
\noindent\textbf{Characteristic Formulas}. In order to formulate our
next axioms
(dealing with special logic gates), we give some characteristic formulas for binary states, considering two qubits indexed by $i$ and $j$:
\smallskip\par\noindent
\begin{tabular}{|l|l|}
\hline
States & Characteristic Formulas\\
 \hline
& \\
$\overline{\mid 00 \rangle_{ij}} = \overline{\diraco_i \otimes \diraco_j}$ & $\langle 0_i ? \rangle 0_j \wedge [1_i ?] \perp$
 \\

 \hline
 Bell states: & \\
 $\beta_{xy}^{i,j} = \overline{\mid 0\rangle_i\otimes \mid y \rangle_j
+ (\mbox{-1})^{x} \mid 1 \rangle_i\otimes \mid \tilde{y}\rangle_j}$ &  $\langle 0_i? \rangle y_j \wedge \langle 1_i ? \rangle \tilde{y}_j \wedge \langle +_i ? \rangle (-)^{x}_j$  \\
 with $\tilde{0} = 1$ and $\tilde{1} = 0$ , $x,y \in \{0,1\}$ & where $(-)^{x} = -$ if $x = 1$ \\
& and $(-)^{x} = +$ if $x = 0$
 \\
 \hline
 $\gamma^{i,j} = \beta_{00}^{i,j} + \beta_{01}^{i,j} =$ & \\
 $ \overline{\mid 00 \rangle_{ij} + \mid 01 \rangle_{ij} + \mid 10 \rangle_{ij} + \mid 11 \rangle_{ij} }$ & $\langle 0_i ? \rangle +_j \wedge \langle 1_i ? \rangle +_j \wedge \langle +_i ? \rangle +_j$\\
 \hline
 \end{tabular}

\bigskip\par\noindent
\textbf{Characteristic Axioms for Quantum Gates $X$ and $Z$.} In general, for all unitary transformations
U 2 U, we have as a consequence of the previous axioms that: $\vdash p_K \to [U_I]p_K$, for $I \cap K = \emptyset$.
\par\noindent
In addition to this, we require for $X,Z,H$:

\begin{eqnarray*}
\begin{array}{lllllllllll}
& \vdash & 0_i \to [X_i]1_i  &;&
& \vdash & 1_i \to [X_i]0_i&;&
& \vdash & +_i \to [X_i]+_i \\
& \vdash & 0_i \to [Z_i]0_i  &;&
& \vdash & 1_i \to [Z_i]1_i  &;&
& \vdash & +_i \to [Z_i]-_i \\
& \vdash & 0_i \to [H_i]+_i  &;&
&  \vdash & 1_i \to [H_i]-_i &;&
&  \vdash &+_i \to [H_i]0_i
\end{array}
\end{eqnarray*}
\smallskip
\noindent\textbf{Notation}. For $x, y\in \{0,1\}$ and distinct indices
$i,j\in N$, we make the following abbreviations for ``Bell formulas'':
$\beta_{xy}^{ij} := \overline{(Z_1^x ; X_1^y)}_{ij}$.

\smallskip\par\noindent
{\bf Propostion 14.} {\it The Bell states  $\beta_{xy}^{i,j}$ are characterized by the
logic Bell
formulas $\beta_{xy}^{ij}$. In other words, a state satisfies one of
these formulas iff it coincides with the corresponding Bell state.}

\smallskip\par\noindent
{\it Proof}: It is enough to check that the formulas $\beta_{xy}^{ij}$
imply the corresponding characteristic formulas in the above
table. For this, we use the Entanglement Axiom and the following (easily checked) theorems:
$\vdash \, 0_1 \leftrightarrow  < Z_1^x; X_1^y > y_1$,
\, $\vdash \, 1_1 \leftrightarrow  < Z_1^x; X_1^y > \tilde{y}_1$,
\par\noindent
$\vdash \, +_1 \rightarrow  < Z_1^x; X_1^y > (-)^x_1$.

\smallskip
\par\noindent
\textbf{Characteristic Axioms for $CNOT$}.
$\, \,$ With the above notations, we put:
\begin{eqnarray*}
\begin{array}{lllllll}
& \vdash & 0_i \wedge c_j \to  [CNOT_{ij}] c_j & ;&
& \vdash & 1_i \wedge 0_j \to   [CNOT_{ij}]1_j \\
& \vdash & 1_i \wedge 1_j \to   [CNOT_{ij}] 0_j &;&
& \vdash &  1_i \wedge +_j \to  [CNOT_{ij}] +_j \\
& \vdash & +_i \wedge 0_j \to [CNOT_{ij}] \beta_{00}^{ij} &;&
& \vdash & +_i \wedge 1_j \to   [CNOT_{ij}]\beta_{01}^{ij} \\
& \vdash & +_i \wedge +_j \to   [CNOT_{ij}]\gamma^{ij}
&{\mbox where } & &\gamma^{ij} &= \langle 0_i ? \rangle +_j \wedge \langle 1_i ? \rangle +_j \wedge \langle +_i ? \rangle +_j
\end{array}
\end{eqnarray*}

\smallskip\par\noindent
{\bf Proposition 15.} {\it
For all $x,y \in \{0,1\}$:
$\vdash \, (H_i ; CNOT_{i,j} (x_i \wedge y_j) = \beta_{xy}^{ij}$
}

\smallskip\noindent\textbf{Corollary}. If $i, j, k$ are all distinct then \\
$ \, \, \vdash (CNOT_{ij}; H_j ; (x_i \wedge y_j) ? ) (p)=_k
\beta_{xy}^{i,j} ? (p)$.

\smallskip\par\noindent
{\it Proof}: From the above and $H^{\dagger} = H$, $CNOT^{\dagger} = CNOT$, we get
$$\vdash \beta_{xy}^{ij} \to [CNOT_{i,j}; H_i] (x_i \wedge y_i)$$
and so
$$\vdash \langle CNOT_{ij}; H_i ; (x_i \wedge y_j) ? \rangle \top \leftrightarrow
\langle \beta_{xy}^{ij} ? \rangle \top $$
The conclusion follows from this, together with the Agreement Axiom.

%%%%%%%%%%%%%%%%%%%%%%%%%%%%%%%%%%%%%%%
\section{Correctness of the Teleportation Protocol}
%%%%%%%%%%%%%%%%%%%%%%%%%%%%%%%%%%%%%%%

Following \cite{Nielsen}, quantum teleportation is the name of a technique that makes it possible to teleport the state of a quantum
system without using a channel that allows for quantum communication,
but with a channel that allows for classical communication.  We are working in $H\otimes H\otimes H$, with $H$ being the
two-dimensional (qubit) space, and so $n=3$.
We assume two agents, Alice and Bob who are separated in space and
each has one qubit of an entangled EPR pair that is represented by
$\beta_{00}^{2,3} \in H^{(2)}\otimes H^{(3)}$.  Alice  holds in addition to her
part of the EPR pair also a qubit $q_1\in H^{(1)}$, in an unknown local state
$q_1$.
Alice ``teleports'' this state to Bob, i.e. she performs a
program that will output a state satisfying $\varphi_3$. To do this, she
first entangles $q_1$ with her part $q_2$ of the EPR pair (i.e. she performs a
$CNOT_{1,2}$ gate on the two qubits and then a Hadamard transformation
$H_1$ on the first component).  Bob's qubit has suffered during the
actions of Alice and when Alice will measure her qubits she will
destroy the entanglement of the EPR pair that she shares with Bob.
The initial state of Bob's qubit is known and we can calculate which
changes it has gone through when we know the result that Alice obtains
from the two measurements.  Moreover, the result that Alice obtains
from the two measurements indicate the actions that Bob has to perform
in order to transfer his qubit $q_3$ into the state $q_1$ was before the protocol.  It is
enough for Alice to send Bob two classical bits encoding the
result $x_1$ of the first measurement and the result $y_2$ of the
second measurement. This means that Bob will have to apply $y$ times
the $X$-gate followed by $x$ times the $Z$ gate, if he wants to
force his qubit $q_3$ into the state $\varphi_3$.
\par\noindent
In our syntax, the quantum program described here is:
$$\pi = \bigcup_{x,y\in \{0,1\}} CNOT_{1 2} ; H_1; (x_1\wedge y_2)?;
X_3^y ; Z_3^x $$
and the validity expressing the correctness of teleportation is
$$\vdash \, \pi (\varphi_1\wedge \beta_{00}^{2,3}) \to [\pi]\varphi_3$$
for all testable $1$-local formulas $\varphi_1$. To show this, observe that by applying the above Corollary (at the end of the last section) in
which we take $i=1, j=2, k=3$ and substitute $p_3$ with $[X_3^y;X_3^x]\varphi_3$, we
we obtain that the validity above (to be proved) is equivalent to:
$\, \, \vdash\, \varphi_1 \wedge \beta_{00}^{23} \to [\beta_{xy}^{1,2}?][X_3^y; Z_3^x]\varphi_3$.
\par\noindent
Replacing the logical Bell formulas with their definitions
$\beta_{xy}^{ij} := \overline{(Z_1^x ; X_1^y)}_{ij}$, we obtain the
following equivalent validity:
$ \vdash \, q_1\wedge \overline{id}_{23} \to [(\overline{ (Z_1^x; X_1^y)_{12}}?][X_3^y; Z_3^x]\varphi_3$, where  $id=Z_1^0; X_1^0$ is the identity. This last validity follows
from applying the Teleportation Property and the validity $\vdash \varphi_1 \to [Z_1^x; X_1^y ; X_1^y; Z_1^x]\varphi_1$ (due to
$X^{\mbox{-1}}=X, Z^{\mbox{-1}}=Z$).
\par\noindent
\textbf{Note}. This proof of correctness can be easily adapted to
cover Logic-Gate Teleportation. Moreover, the whole range of quantum
programs covered by the ``entanglement networks'' in \cite{Coecke2} can
be similarly treated using our logic.

\small{

}

\end{document}